\documentclass[prd,superscriptaddress]{revtex4}
\usepackage{graphicx}
\usepackage{epsf}
\usepackage{epsfig}
\usepackage{amssymb}
\usepackage{amsmath}

\def\1{{\'{\i}}}

\newcommand{\be}{\begin{equation}}
\newcommand{\ee}{\end{equation}}
\newcommand{\bea}{\begin{eqnarray*}}
\newcommand{\eea}{\end{eqnarray*}}

\begin{document}
\title{Exploring the R\'enyi Holographic Dark Energy Model with the Future and the Particle Horizons as the Infrared Cut-off}
\author{Suphakorn Chunlen\footnote{E-mail address: suphakorn.official@gmail.com}}
\affiliation{Department of Physics, School of Science, University of Phayao, 19 Moo 2, Mae Ka, Muang, Phayao 56000, Thailand.}
\author{Phongsaphat Rangdee\footnote{E-mail address: phongsaphat.ra@up.ac.th}}
\affiliation{Department of Physics, School of Science, University of Phayao, 19 Moo 2, Mae Ka, Muang, Phayao 56000, Thailand.}

\begin{abstract}
We study the R\'enyi holographic dark energy (RHDE) model by using the future and the particle horizons as the infrared (IR) cut-off. With the initial condition from the literature, most of the cosmological parameters are computed. Some of the results agree with the observation that the present universe is in accelerating expansion and in a phantom phase.
\end{abstract}

\pacs{95.36.+x; 98.80.-k}

\maketitle

\section{Introduction}
It is widely known that the universe is expanding with acceleration \cite{Riess:1998cb,Hinshaw:2012aka,Aghanim:2018eyx}. Many theoretical models have been constructed to explain this behaviour. One of them are the dark energy model. There are various types of the dark energy model. The most common and acceptable one is the Lambda Cold Dark Matter ($\Lambda$CDM) model. $\Lambda$CDM has given consistent results with the observation, but it suffers from the cosmological constant problem \cite{Peebles:2002gy,Padmanabhan:2002ji,Copeland:2006wr,Frieman:2008sn,Bamba:2012cp,Wang:2016och}. A lot of new dark energy models have been established to solve this issue. One of those is the holographic dark energy (HDE) model proposed in \cite{Li:2004rb}. Inspired by the holographic principle \cite{tHooft:1993dmi}, the HDE model has the dark energy density depending on the universe boundary physical quantities i.e. the reduced Planck mass $M_p$ and the cosmological length scale or the IR cut-off $L$. Several choices of the IR cut-off have been considered \cite{Li:2004rb,Cruz:2018lcx,Sharma:2020glf}, such as the Hubble horizon, the future horizon and the particle horizon, and in addition some new IR cut-offs have been proposed such as Nojiri-Odintsov cut-off \cite{Nojiri:2005pu} and Granda-Oliveros one \cite{Granda:2008dk}. The energy density of the original HDE model has arisen from the black hole entropy or the Bekenstein-Hawking entropy \cite{Li:2004rb,Horava:2000tb}. Some literature also has the studies of the HDE model with modified gravity theories \cite{Nojiri:2017opc,Setare:2009ym,Chattopadhyay:2012eu,Farooq:2013ava,Pasqua:2014owa,Granda:2019agf,Kritpetch:2020vea} and additionally the model unified with the inflation regime \cite{Nojiri:2019kkp,Nojiri:2020wmh}. The original HDE with the Hubble horizon as the IR cut-off cannot lead to the accelerating expansion of the present universe \cite{Li:2004rb}. As a cure, using the future horizon yields the acceleration, however, it gives the big rip singularity \cite{Li:2004rb}. By using generalized entropies for further investigation, many types of HDE models are formed, such as R\'enyi HDE \cite{Moradpour:2018ivi}, Tsailis HDE \cite{Tavayef:2018xwx} and Sharmal-Mittal HDE \cite{Jahromi:2018xxh}. For those models, different aspects have been investigated e.g. in \cite{Nojiri:2019skr,Sharma:2020wba,Dubey:2020ckn,Golanbari:2020coz,Iqbal:2019ooy,Moradpour:2019jrp,Sharif:2019seo,Ghaffari:2018rzs,Saridakis:2018unr}, and also there are some papers that the (dark) matter and the dark energy interaction has been considered such as \cite{Sharma:2019bgp,Saha:2020vxn,Sharma:2020glf,Bhattacharjee:2020rqk,Jawad:2018juh,Mamon:2020ocb,Aditya:2019bbk,Ghaffari:2019qcv,Zadeh:2019yfy,Ghaffari:2018wks,Zadeh:2018poj}. The R\'enyi HDE model with the Hubble horizon is stable without including any interaction \cite{Moradpour:2018ivi} where the universe at present is in the quintessence phase. The recent observation \cite{Amanullah:2010vv,Hinshaw:2012aka,Tripathi:2016slv,Aghanim:2018eyx,Tripathy:2019jsk} suggested the universe phantom phase is favoured at the present time, so considering other regimes of the R\'enyi HDE model without interaction would yield the present phantom phase corresponding to the observation. In this research, we investigate the R\'enyi HDE model in two cases without considering the interaction. The first case is studied by using the future event horizon as the IR cut-off, and we use the particle horizon as the cut-off for the second case. The paper is organised as follows. First, we introduce a brief formation of the R\'enyi HDE from the literature. Next, we derive equations describing the dynamics of the model where some of the cosmological parameters are presented. Then we show the results containing some cosmological parameters i.e. the deceleration parameter, the equation of state parameter, the dark energy density parameter, the ratio between the (dark) matter and the dark energy, and the square of the sound speed of the dark energy. Finally, we make a conclusion.

\section{The R\'enyi Holographic Dark Energy in Brief}
From \cite{Moradpour:2018ivi}, in the unit of $c=G=\hbar=k_B=1$, the R\'enyi entropy \cite{RenyiProbTheory1970} is related to the Tsailis entropy \cite{Tsallis:1987eu} by \cite{Biro:2011:APSPhysRevE.83.061147,Czinner:2015eyk,Moradpour:2018ivi,Komatsu:2016vof,Moradpour:2019jrp,Moradpour:2018ima}
\begin{equation}
	S_R=\frac{1}{\delta}\ln(1+\delta S_T), \label{eq:RenyiEntropy}
\end{equation}
where $S_R,S_T,\delta$ are the R\'enyi entropy, the Tsailis entropy, and a real parameter respectively. The Bekenstein-Hawking entropy is a type of the Tsailis entropy \cite{Majhi:2017zao,Biro:2013cra,Moradpour:2018ivi}, so substituting this into \eqref{eq:RenyiEntropy} yields \cite{Moradpour:2018ivi}
\begin{equation}
	S_R=\frac{1}{\delta}\ln(1+\delta \frac{A}{4}),
\end{equation}
where $A$ is the surface area of the black hole horizon. For the case of holographic dark energy, $A=4\pi L^2$ where $L$ is the cosmological length scale or the IR cut-off. From the second law of thermodynamics, the assumption $\rho_DdV\propto TdS_R$ \cite{Moradpour:2018ivi} leads to
\begin{equation}
	\rho_D=\frac{3C^2}{8\pi L^2(1+\delta\pi L^2)}, \label{eq:RenyiEnergyDensity}
\end{equation}
which is the R\'enyi holographic dark energy density where $C$ is a numerical constant \cite{Li:2004rb,Moradpour:2018ivi}.

\section{Equations for Describing the Universe Dynamics}
We work in the flat Friedmann--Lema\^\i{}tre--Robertson--Walker spacetime with the metric
\begin{equation}
	ds^2=-dt^2+a^2(t)\left(dr^2+r^2d\Omega_2^2\right).
\end{equation}
Applying this together with the matter and the dark energy constituents \cite{Genova-Santos:2020tfc} to the Einstein field equations, we have the Friedmann equations:
\begin{equation}
	H^2=\frac{8\pi}{3}(\rho_D+\rho_m)\qquad\text{or}\qquad1=\Omega_D+\Omega_m \label{eq:Friedmann1}
\end{equation}
and
\begin{equation}
	2\dot H+3H^2=-8\pi P_D, \label{eq:Friedmann2}
\end{equation}
where $H=\dot a/a$, $\rho_D$, $\rho_m$, $\Omega_D=8\pi\rho_D/\left(3H^2\right)$, $\Omega_m=8\pi\rho_m/\left(3H^2\right)$, and $P_D$ are the Hubble parameter, the dark energy density, the matter density, the dark energy density parameter, the matter density parameter, and the pressure of dark energy, respectively. Additionally, from the energy-momentum conservation, the continuity equations are given by
\begin{equation}
	\dot\rho_m+3H\rho_m=0, \label{eq:fluidrhom}
\end{equation}
\begin{equation}
	\dot\rho_D+3H(\rho_D+P_D)=0. \label{eq:fluidrhoD}
\end{equation}
Obviously, the equation \eqref{eq:fluidrhom} yields $\rho_m=\rho_{m0}a^{-3}$ where $\rho_{m0}$ is the matter density at present. Now we consider two cases of the IR cut-off i.e. the future horizon \cite{Li:2004rb}
\begin{equation}
	L_f=a(t)\int_t^\infty\frac{dt'}{a(t')}=\frac{a}{H_0}\int_a^\infty\frac{da'}{a'^2E(a')} \label{eq:futurehorizon}
\end{equation}
and the particle horizon \cite{Li:2004rb,Fischler:1998st}
\begin{equation}
	L_p=a(t)\int_0^t\frac{dt'}{a(t')}=\frac{a}{H_0}\int_0^a\frac{da'}{a'^2E(a')}, \label{eq:particlehorizon}
\end{equation}
where in the latter equalities $H_0$ is the present value of the Hubble parameter $H$ and for later convenience the definition $E(a)\equiv H/H_0$ is used. Let us consider the R\'enyi holographic dark energy density. From \eqref{eq:RenyiEnergyDensity}, we solve for $L^2$ as
\begin{equation}
	L^2=-\frac{1}{2\delta\pi}\pm\sqrt{\frac{1}{(2\delta\pi)^2}+\frac{3C^2}{8\pi^2\delta\rho_D}}.
\end{equation}
For the case of $\delta>0$, it is obvious that we have to choose the plus sign, and for $\delta<0$ both the plus and the minus signs look eligible. However, we encountered a numerical instability when we chose the plus sign for $\delta<0$, so we use the minus sign for it. Using the initial condition at present $a=a_0=1$, from the energy density \eqref{eq:RenyiEnergyDensity} and the Friedmann equation \eqref{eq:Friedmann1}, we are able to determine the numerical constant
\begin{equation}
	C^2=H_0^2(1-\Omega_{m0})L_0^2(1+\delta\pi L_0^2),
\end{equation}
where $L_0$ is the cut-off at $a=1$. Using this together with $H(a)=E(a)H_0$, the Friedmann equation \eqref{eq:Friedmann1} and the energy density \eqref{eq:RenyiEnergyDensity}, we have an equation for $E(a)$ as
\begin{equation}
	E^2(a)=\Omega_{m0}a^{-3}+\frac{(1-\Omega_{m0})L_0^2(1+\delta\pi L_0^2)}{L^2(1+\delta\pi L^2)}. \label{eq:inteqofE}
\end{equation}
To obtain the universe evolution, the function $E(a)$ must be known. We can substitute the future or the particle horizons into the above equation, and we have a nonlinear integral equation of $E(a)$ which is quite tricky to solve numerically. An easier way is to attain its differential version and solve it. We do this by differentiating \eqref{eq:futurehorizon} and \eqref{eq:particlehorizon} with respect to the scale factor $a$, so we get
\begin{equation}
	\frac{d}{da}\left(\frac{LH_0}{a}\right)=\frac{\theta}{a^2E(a)} \label{eq:eomofE}
\end{equation}
where $L$ obtained from \eqref{eq:inteqofE} is
\begin{equation}
	L=\sqrt{-\frac{1}{2\delta\pi}\pm\sqrt{\frac{1}{4\delta^2\pi^2}+\frac{C^2}{\delta\pi H_0^2(E^2(a)-\Omega_{m0}a^{-3})}}}.
\end{equation}
Again the \{plus, minus\} signs are for \{$\delta>0$, $\delta<0$\}. In \eqref{eq:eomofE}, we have $\theta\in\{-1,+1\}$ corresponding to the \{future, particle\} horizon cases. We eventually have \eqref{eq:eomofE} as the differential equation of motion for $E(a)$, and we then solve it numerically for the dynamics of the universe. The cosmological parameters can be written in terms of $E(a)$, the scale facter $a$, and the initial condition values. Some of them are the deceleration parameter
\begin{equation}
	q\equiv -1-\frac{\dot H}{H^2}=-1-\frac{a}{E(a)}\frac{dE(a)}{da},
\end{equation}
the dark energy pressure
\begin{equation}
	p_D=-\frac{H_0^2E(a)}{8\pi}\left[2aE'(a)+3E(a)\right],
\end{equation}
the dark energy density
\begin{equation}
	\rho_D=\frac{3H_0^2}{8\pi}\left[E^2(a)-\frac{\Omega_{m0}}{a^3}\right],
\end{equation}
the total equation of state parameter $w=p_D/\left(\rho_D+\rho_m\right)$ \cite{Moradpour:2018ivi}, and for instability investigation the square of the sound speed $v_s^2=dp_D/d\rho_D$. The results are shown in the next section. 

\section{Results and Discussion}
\begin{figure}[h]
	\begin{center}
		\includegraphics[width=.48\textwidth]{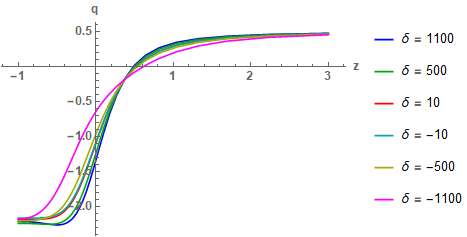}\quad
		\includegraphics[width=.48\textwidth]{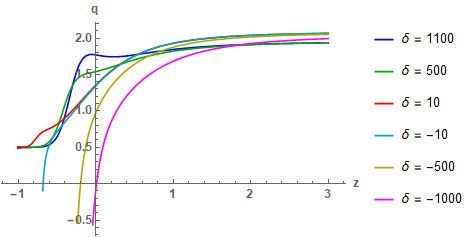} 
	\end{center}
	\caption{Plots of the deceleration parameter $q$ vs the redshift parameter $z$ for different values of $\delta$ where the left and the right graphs are for the future and the particle horizon cases respectively.}
	\label{fig:qzplots}
\end{figure}

\begin{figure}[h]
	\begin{center}
		\includegraphics[width=.48\textwidth]{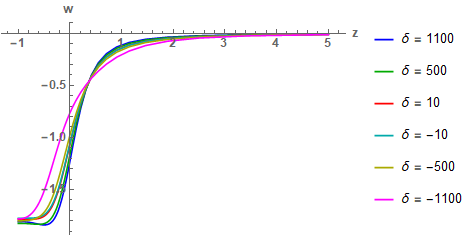}\quad
		\includegraphics[width=.48\textwidth]{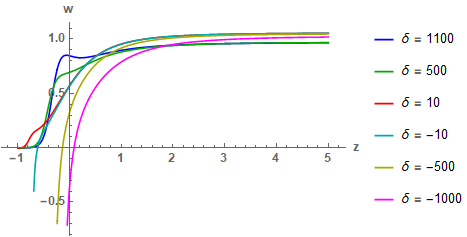} 
	\end{center}
	\caption{Plots of the cosmic equation of state parameter $w$ vs the redshift parameter $z$. The left and the right plots are for the future and the particle horizon cases respectively.\\}
	\label{fig:wzplots}
\end{figure}

\begin{figure}[h]
	\begin{center}
		\includegraphics[width=.48\textwidth]{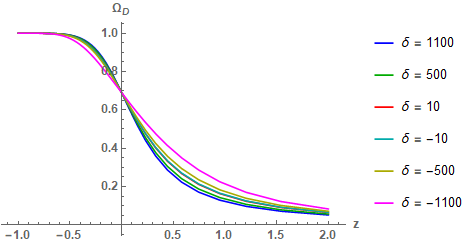}\quad
		\includegraphics[width=.48\textwidth]{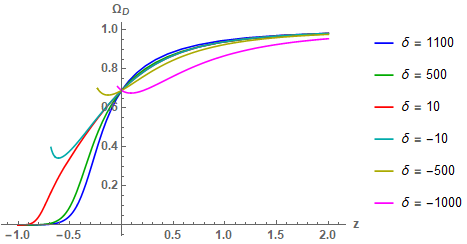} 
	\end{center}
	\caption{Plots of the dark energy density parameter $\Omega_D$ vs the redshift parameter $z$ where the left and the right plots are for the future and the particle horizon cases respectively.}
	\label{fig:omegazplots}
\end{figure}

\begin{figure}[h]
	\begin{center}
		\includegraphics[width=.48\textwidth]{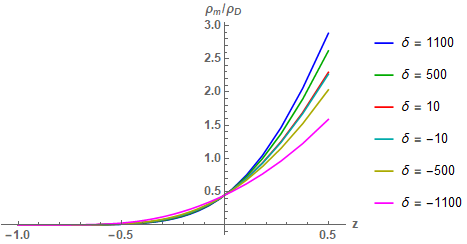}\quad
		\includegraphics[width=.48\textwidth]{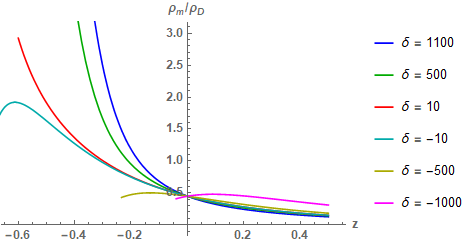} 
	\end{center}
	\caption{Plots of the ratio between the (dark) matter and the dark energy density $\rho_m/\rho_D$ vs the redshift parameter $z$. The left and the right plots are for the future and the particle horizon cases respectively.}
	\label{fig:rhomrhoDzplots}
\end{figure}

\begin{figure}[h]
	\begin{center}
		\includegraphics[width=.48\textwidth]{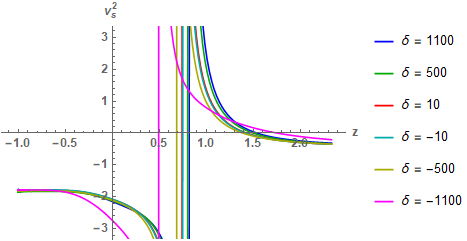}\quad
		\includegraphics[width=.48\textwidth]{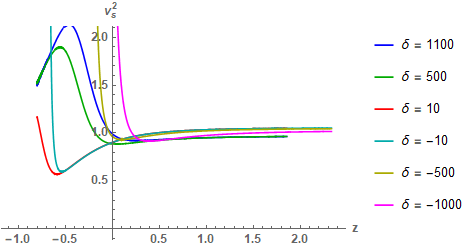} 
	\end{center}
	\caption{Plots of the square of the dark energy sound speed $v_s^2$ vs the redshift parameter $z$. The left plot is for the future horizon case while the right plot is for the particle horizon one.}
	\label{fig:vssqzplots}
\end{figure}

We refer to the initial data from the Planck 2018 results \cite{Aghanim:2018eyx}. In particular, we use $H_0=67$ km/s/Mpc and $\Omega_{m0}=0.31$ for the present values of the Hubble parameter and the matter density parameter respectively. Graphs of the deceleration parameter as a function of the redshift parameter $z=1/a-1$ for the cases of the future and the particle horizons are shown in figure \ref{fig:qzplots}.
From the plots, the future horizon case gives the present accelerating expansion of the universe as well as the transition from a decelerating phase to an accelerating one. On the other hand, the particle horizon case, for $\delta>0$, yields only decelerating expansions of the universe, but for $\delta<0$ it has the transition between the decelerating expansion to the accelerating one, despite its small negative value of $q$ at present for the case of $\delta=-1000$. Additionally, the plots for $\delta<0$ in the particle horizon case end before reaching $z=-1$ due to a numerical instability. The graphs of the cosmic equation of state parameter $w$ are shown in figure \ref{fig:wzplots}.
In the future horizon case, for most of the $\delta$ values, at present time the universe goes through the phantom phase which has been favoured by the recent observation \cite{Amanullah:2010vv,Hinshaw:2012aka,Tripathi:2016slv,Aghanim:2018eyx,Tripathy:2019jsk}. In the case of the particle horizon, for $\delta>0$ the cosmic equation of state is always not negative. This means the universe is always in a decelerating phase. For $\delta<0$ the equation of state can go less than zero. This indicates the quintessence phase \cite{Goswami:2019zto} of the universe. Moreover, the dark energy density parameter $\Omega_D$ and the ratio between the (dark) matter and the dark energy, $\rho_m/\rho_D$, are plotted in figure \ref{fig:omegazplots} and \ref{fig:rhomrhoDzplots} respectively.
In figure \ref{fig:omegazplots} and \ref{fig:rhomrhoDzplots}, for the future horizon case, the universe is in the matter-dominated era in the beginning $z>0$ and then turns to the dark energy dominated era at present which corresponds to the $\Lambda$CDM model. However, for the case of the particle horizon the universe starts with the dark energy dominated era and for $\delta>0$ it turns to matter-dominated era in the cosmic future. This contradicts to the standard cosmology. For instability analysis, let us consider the sound speed squared of the dark energy. The plots of it are in figure \ref{fig:vssqzplots}.
At present $z=0$, we have $v_s^2<0$ for the future horizon case. This means the future horizon choice of the IR cut-off for the RHDE model yields an unstable universe, though we have not yet considered the interaction between the (dark) matter and the dark energy. In the particle horizon case, it appears to be stable at present for most of the $\delta$ values considered.

\section{Conclusion}
We have investigated the R\'enyi HDE model of which the IR cut-off is the future and the particle horizons. For the future horizon case, it shows that the present universe is in the accelerating expansion and in the phantom phase. It also shows the transition from deceleration to acceleration of the expansion in the cosmic past. However, there is instability occurred due to the negativity of the square of the sound speed, $v_s^2<0$. The case of the particle horizon, on the other hand, has two different subcases showing different results. The subcase of $\delta>0$ does not yield the acceleration of the universe expansion at any point in cosmic history. Furthermore, the subcase of $\delta<0$ can give either the present acceleration or deceleration of the universe expansion. This subcase is stable at present for some values of $\delta$ since the sound speed squared is not negative. However, the matter constituent history for this subcase does not correspond to $\Lambda$CDM that the (dark) matter is dominated in the past of cosmic time. We hope inclusion of interaction between the (dark) matter and the dark energy will alleviate the domination and the instability issues which are left for future work.

\acknowledgments{This research was supported by the Thai government budget grant RD62040, University of Phayao. We thank Daris Samart and Chakkrit Kaeonikhom for useful discussion.}


\end{document}